# Résolution des conflits sémantiques pour l'intégration des composants métier.

**Larbi KZAZ\* -Hicham ELASRI\*\* -Abderrahim SEKKAKI\*\***

*\*Institut Supérieur de Commerce et d'Administration des Entreprises (ISCAE)*
 *Km 9,5 Route de Nouasseur BP. 8114 - Casablanca Oasis*
*Tél. : + 212 5 22 33 54 82 à 85  Fax : +212  522 33 54 96*
*Kzaz_larbi@yahoo.fr*

*\*\* Laboratoire d'Informatique et d'Aide á la Décision (LIAD)*
*Université Hassan II  Faculté des Sciences Ain Chock Casablanca Maroc*
*B.P 5366 Maarif Casablanca 20100 Maroc*
*Tél : +212 522 23 06 80 / 84. Fax : +212 522 23 06 74.*
*hicham_elasri@yahoo.com a_sekkaki@yahoo.fr*

RÉSUMÉ. *La réutilisation et l'intégration des Composants Métier dans un nouveau SI exige la détection et la résolution des conflits sémantiques. Par ailleurs les systèmes d'intégration et de résolution des conflits sémantiques se basent la plupart du temps sur des méthodes d'alignement des ontologies basées sur une ontologie de support. Ce travail se positionne à l'intersection de ces deux thématiques de recherche : Intégration des CM en vue de leur réutilisation et alignement des ontologies en vue de la résolution des conflits sémantiques. Notre contribution  concerne aussi bien la proposition d'une solution pour l'intégration des CM fondée sur l'alignement des ontologies, qu'une méthode d'enrichissement de l'ontologie de domaine choisie comme ontologie de support.*

ABSTRACT. *Reusing and integrating Business Components in a new Information System requires detection and resolution of semantic conflicts. Moreover, most of integration and semantic conflict resolution systems rely on ontology alignment methods based on domain ontology. This work is positioned at the intersection of two research areas: Integration of reusable B C  and alignment of ontologies for semantic conflict resolution. Our contribution concerns both the proposal of a BC integration solution based on ontologies alignment and a method for enriching the domain ontology used as a support for alignment.*

MOTS CLES: *Composants Métier, Intégration sémantique, alignement des ontologies, Enrichissement des Ontologies.*

## 1. Introduction

L'approche à base de composants est considérée depuis les années 90, comme un nouveau paradigme de développement des systèmes d'information (Barbier, 2002). Elle a pour but de réduire de façon conséquente les coûts et les délais de mise à disposition des logiciels. Le principe de cette approche consiste à construire de nouveaux systèmes à partir de composants disponibles. Le recours à cette approche vise aussi bien les phases de développement que les phases amont d'analyse et de conception, où les composants ne sont plus techniques mais des modèles qui capturent un certain savoir sur un domaine. Les composants concernés dans les phases d'analyse et de conception sont communément connues sous le terme de Composant Métier (CM) ou Business Component. De nombreux travaux se sont intéressés à la conception de nouveaux SI à partir de composants réutilisables (Saidi, 2009), (Barbier et al., 2002).Ces travaux ont mis en évidence plusieurs problématiques dont la recherche, l'adaptation, la composition et l'intégration des CM. Nous nous intéressons dans ce travail à la question de l'intégration des CM dans un même SI. En effet, l'intégration de CM issus de différentes sources engendre plusieurs types de conflits structurels, syntaxiques, sémantiques etc. Nous nous focalisons sur la détection et la résolution des conflits sémantiques de type nommage, rencontrés lors du processus d'intégration des composants métier (Kzaz et al., 2010). Par ailleurs les systèmes d'intégration sémantique se basent la plupart du temps sur des méthodes d'alignement d'ontologies (Safar et al., 2009). Nous nous appuyons sur les résultats de certains travaux sur l'alignement pour supporter le processus d'intégration sémantique des CM.

Notre papier est organisé comme suit : Dans la section 2 nous présentons la problématique de l'intégration sémantique des CM et les différents conflits qu'elle soulève. Dans la section suivante nous proposons un processus d'intégration sémantique des CM, fondé sur l'alignement et l'enrichissement des ontologies ainsi qu'une méthode de mesure pour la détection et la résolution des conflits sémantiques de type nommage. La section 4 décrit la méthode d'enrichissement de l'ontologie de domaine préconisée. La section 5 présente brièvement le prototype de validation du processus. En fin d'article nous exposons les différentes perspectives d'extension de notre solution.

## 2. Intégration sémantique des Composants Métier.

Le terme composant est largement utilisé dans le domaine de la réutilisation, avec une connotation générale d'entité autonome réutilisable. Dans la littérature, on trouve plusieurs définitions du concept de composant métier (Herzum, 2000), (Barbier, 2002). Ces définitions considèrent qu'un CM modélise et implémente une entité significative par rapport à un métier de l'entreprise. Il capture et décrit dans des termes issus du vocabulaire de l'entreprise et de son métier, des concepts, des événements ou des processus. Un CM peut modéliser aussi bien une entité qu'un processus de l'entreprise. Il peut être aussi bien un modèle statique, dynamique ou fonctionnel. Selon l'approche à base de CM, un SI d'entreprise est construit à partir d'un ensemble de CM pouvant émaner de différentes sources. Le SI commercial d'une entreprise, par exemple, pourrait être conçu à partir de CM tels que : {«Processus de Vente », « Produit », « Commande » etc..}.

L'intégration dans un même SI de plusieurs CM, issus de différentes sources, nécessite la détection et la résolution de conflits structurels, terminologiques, syntaxiques et sémantiques. Plusieurs chercheurs, (Hendriks, 2007) ,(Kavouras , 2004) ,(Izza, 2006), (Visser, 2001),ont identifié trois types de conflits sémantiques: conflit de confusion, conflit de mesure et conflit de nommage. (Visser, 2001) définit le conflit de type nommage, auquel nous nous limitons, comme suit : *«Naming conflicts occurs when naming schemes of information differ significantly. A frequent phenomenon is the presence of homonyms and synonyms »*.

L'exemple suivant (figure 1), illustre les conflits de nommage entre les concepts présents dans deux CM. Les concepts de « *Service* » dans CM1 et dans CM2 sont homonymes ; ils utilisent le même terme pour désigner des concepts représentant des entités différentes. Alors que le concept de « *Service* » dans CM1 est synonyme avec le concept de « *Prestation* » dans CM2. De même les concepts de « *Compagnie* » et de « *Cabine*t » sont des synonymes.

L'intégration de ces deux composants métier devrait aboutir à l'obtention d'un nouveau Composant regroupant les concepts des deux CM (figure 1.).

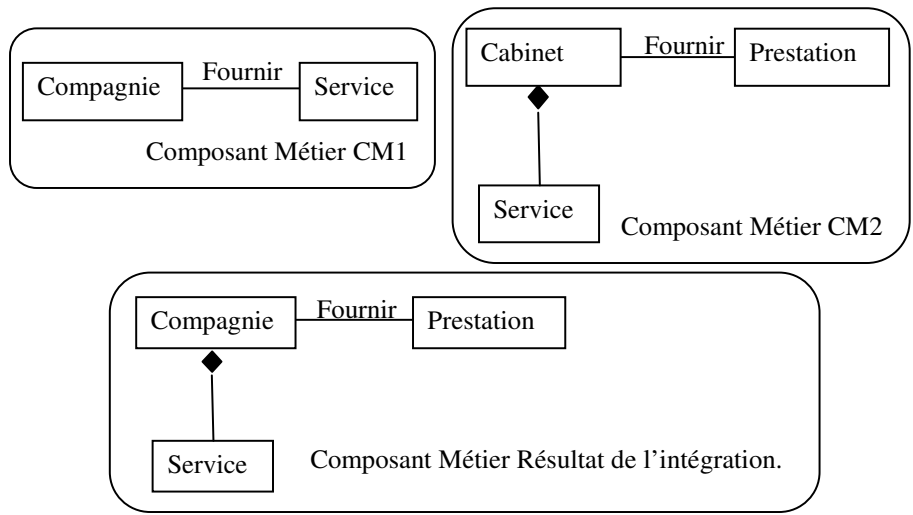

Figure 1. Intégration des CM après détection et résolution des conflits sémantiques..

Dans la littérature on peut distinguer deux catégories de mécanismes d'intégration : les mécanismes qui se basent sur des modèles préétablies (les modèles de composant) et les mécanismes qui se basent sur les ontologies. Notre proposition s'appuie sur les ontologies, comme élément clé pour assurer la résolution des conflits.

En effet un CM modélise des entités et des processus d'un domaine de gestion dans des termes librement choisis par les concepteurs. Les ontologies de domaine quant à elles, sont des conceptualisations des objets reconnus comme existant dans un domaine, de leurs propriétés et des relations les reliant (Fürst, 2004). Les concepts et les relations présentes dans une ontologie sont supposés faire l'objet d'un consensus entre les experts du domaine. Le passage par une ontologie de domaine permettra donc de résoudre les conflits sémantiques de nommage.

## 3. Processus d'intégration sémantique des Composants Métier.

Le processus d'intégration sémantique que nous proposons permet de :

- Détecter et de résoudre les conflits sémantiques de type nommage entre les composants métier candidats à une intégration dans le nouveau SI.
- Produire un nouveau CM résultant de l'intégration des composants métier de départ.
- Enrichir l'ontologie de domaine utilisée comme ontologie de support lors du processus d'intégration.

Notre proposition s'appuie sur les résultats de plusieurs travaux de recherche, notamment ceux concernant la transformation des CM décrits dans des langages de modélisation en langages de description d'Ontologies (Faucher et al., 2006), (Gašević et al., 2004b) et ceux relatifs à l'alignement et à l'enrichissement des ontologies de domaine (Sabou et al., 2006), (Safar et al., 2009), (Aleksovski et al., 2006), (Di Jorio et al., 2007), (Ben Ghezaiel et al., 2010), (Faucher et al., 2006), (Mhiri et al., 2006). La figure 2 fournit une représentation en boîte noire du processus.

Les Entrées du processus d'intégration sont :

- Un ensemble de Composants Métier, notés CM1,… CMn, sélectionnés par le concepteur en vue de leur intégration dans le futur SI.
- Une Ontologie de Domaine choisie par le concepteur en fonction du domaine dont relève le SI à construire ; elle servira par la suite d'ontologie de support au cours du processus d'intégration.

Le processus d'intégration fournit en sortie :

- Un nouveau CM résultat de l'intégration de l'ensemble des CM de départ.
- Une nouvelle ontologie de domaine enrichie par de nouvelles relations sémantiques ajoutées suite à deux traitements que nous appliquons dans le processus : alignement et enrichissement des ontologies.

Les deux sorties du processus peuvent être réutilisées ultérieurement dans de futures intégrations concernant de nouveaux composants :

- Le nouveau CM résultat peut être utilisé en tant que candidat à l'intégration avec d'autres composants.
- L'ontologie de domaine enrichie peut être utilisée pour mettre à jour l'ontologie de domaine initiale et servir comme nouvelle ontologie de support dans de futures itérations d'intégration permettant ainsi d'augmenter l'efficacité du processus.

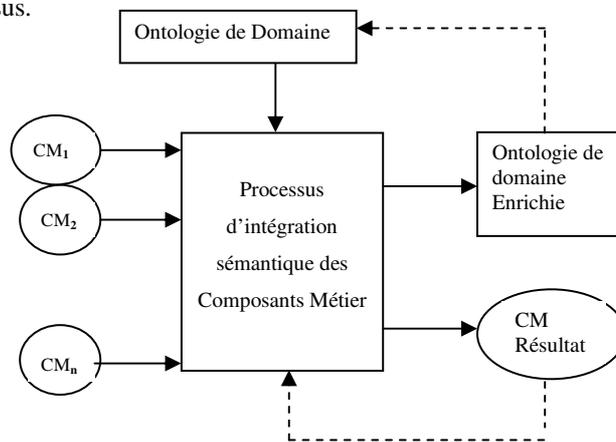

Figure 2 : Représentation en boite noire du processus.

Le processus d'intégration se déroule en trois étapes :

1. Transformation des CM candidats à l'intégration en ontologies.
2. Alignement des ontologies obtenues en s'appuyant sur une ontologie de domaine.
3. Transformation de l'ontologie résultat de l'alignement en un nouveau Composant Métier intégrant l'ensemble des CM source.

*3.1 Transformation des CM candidats à l'intégration en ontologies.*

Les CM candidats à l'intégration sont des représentations conceptuelles d'entités et de processus d'un domaine d'entreprise; ils sont décrits dans un langage (formel, semi formel ou naturel) de modélisation donné. Plusieurs travaux de recherche se sont intéressés récemment à la transformation des modèles conceptuels décrits dans un langage tel qu'UML en modèles utilisant les langages de description des ontologies (Faucher et al., 2006), (Gašević et al., 2004b). Nous nous basons sur ces travaux pour assurer la transformation des CM en ontologies.

*3.2 Alignement des ontologies obtenues en s'appuyant sur une ontologie de domaine.*

L'alignement des ontologies vise à établir des correspondances entre deux ontologies, portant à priori sur le même domaine de connaissance. Il consiste à trouver des relations sémantiques entre des concepts définis dans les ontologies à aligner. Afin de réaliser l'intégration sémantique des CM, nous nous appuyons sur une méthode d'alignement des ontologies basée sur une ontologie de support, appelée aussi dans la littérature (Safar et al., 2009) ontologie de background. L'ontologie de support est dans notre cas, l'ontologie relative au domaine dont relève le SI à construire. Cette étape du processus prend en entrée :

- Les ontologies obtenues dans l'étape précédente.
- L'ontologie de domaine.

Elle fournit en résultat :

- Une ontologie résultat de l'alignement de l'ensemble des ontologies fournies en entrée.
- L'ontologie de domaine enrichie par de nouvelles relations.

Afin de réaliser l'alignement des ontologies relatives aux différents CM nous proposons une méthode de mesure de similarité sémantique, notée $\sigma$ par la suite, s'appuyant sur une méthode de mesure de la similarité syntaxique notée $\sigma'$ ainsi que sur une ontologie de domaine notée $O_d$.

Soit $E_{ci}$ l'ensemble des concepts présents dans l'ontologie de composant $OCM_i$.

Soit $E_c$ l'ensemble des concepts présents dans l'ensemble des ontologies de composants : $E_c = Union (EC_i)$ $1<=i<=n$.

Soient C1, C2 deux concepts appartenant à $E_c$.

Soit Terme(Ci) est une fonction qui retourne le terme utilisé pour désigner le concept Ci.

*Mesure de la similarité syntaxique*

La méthode de mesure de similarité syntaxique que nous proposons est notée σ' ; elle prend la valeur 1 lorsque les concepts sont syntaxiquement identiques et 0 dans le cas contraire. Elle est décrite dans l'encadré n° 1 ci-après.

---

**σ' : Ec ×Ec → {0, 1}**
**Entrées :** Les deux concepts C1 et C2 à comparer syntaxiquement.
**Sorties : 1** si C1 et C2 sont syntaxiquement identiques ; **0** sinon.
**Début.**
 **Si** C1 et C2 sont des concepts atomiques **alors**
    **Si** *Terme(C1) =Terme(C2)* **Alors**   *σ' (C1, C2) = 1*
    **Sinon**                              *σ' (C1, C2) = 0*
    **Finsi**
 **Sinon**
        % C1 et C2 sont composites. C1 et C2 s'écrivent alors C1 = (C11..,  C1i,….., C1n) et C2 = (C21 …., C2j,…., C2n) %
        *σ' (C1, C2) =1/n (Σi j σ' (C1i, C2j)) 1 <= i, j <=n*
 **Finsi**
**Fin**

**Encadré n° 1 :** Méthode de mesure de la similarité sémantique

---

*Mesure de la similarité sémantique.*

La méthode de mesure de la similarité sémantique entre concepts, est basée sur l'ontologie de domaine et sur la méthode de mesure de la similarité syntaxique σ', définie ci-avant.

Soient C1 et C2 deux concepts de Ec, Od l'ontologie de domaine, Rod l'ensemble des relations sémantiques disponibles dans Od, et R(C1,C2) le sous ensemble des relations existantes entre les concepts C1 et C2 au sein de Od. Rod contient R (C1, C2.)

σ, la méthode de calcul de la similarité sémantique est définie dans l'encadré n° 2 ci-après.

```
σ : Ec × Ec → {0, 1},

Entrées :   - Les deux concepts C1 et C2 à comparer sémantiquement
            - Od l'ontologie de domaine
Sorties: 1 s'il existe une relation sémantique de synonymie entre C1 et C2 ;  0 sinon.
Début
  Si (C1 et C2 appartiennent à OD) alors
        Si (R (C1, C2) est vide) alors
              Lancer le traitement d'enrichissement de l'ontologie
              Si une nouvelle relation a été détectée alors
                    Mettre à jour Rod et R (C1, C2)
                    Relancer σ (C1, C2)
              Sinon % La mesure de la similarité sémantique
                    coïncide avec la mesure de similarité syntaxique %
                    σ (C1, C2)= σ' (C1, C2)
              Finsi
        Sinon
              Si R (C1, C2) contient une relation de synonymie alors
                    σ (C1, C2) = 1
              Sinon
                 Si R (C1, C2) Contient une relation d'homonymie alors
                    σ (C1, C2) = 0
                 Sinon
                    σ (C1, C2)= σ' (C1, C2)
                 Finsi.
              Finsi.
        Finsi
  Sinon
        % C1 ou C2 n'appartiennent pas à OD, La mesure de la similarité
        sémantique coïncide avec la mesure de similarité syntaxique %
        σ (C1, C2)= σ' (C1, C2)
  Finsi.
Fin
```

**Encadré n° 2 :** Méthode de mesure de la similarité sémantique

### 3.3 Transformation de l'ontologie résultat de l'alignement en un nouveau Composant Métier intégrant l'ensemble des CM sources.

Il s'agit de la dernière étape du processus d'intégration ; elle consiste simplement à transformer l'ontologie OCMr résultat de l'alignement de l'ensemble des OCMi en

Composant Métier résultat de l'intégration de l'ensemble des CM de départ. Le composant métier obtenu pouvant être réutilisé comme entrée du processus afin de l'intégrer avec d'autres CM.

D'après les travaux de (Gašević et al., 2004), (Knublauch, 2003) et (Grønmo et al., 2005), il s'avère possible de transformer des ontologies vers UML. L'ontologie résultat OCMr est par conséquent transformable en un composant métier représenté en UML.

A cette étape du processus d'intégration deux voies se présentent :

- Transformation automatique de l'ontologie résultat (OCMr) en un composant résultat CMr qui sera par la suite intégré dans le futur SI.
- Les concepteurs pourront utiliser l'ontologie résultat dans une première phase pour détecter et résoudre les conflits sémantiques et dans une deuxième phase pour transformer manuellement l'OCMr en un composant métier résultat CMr à intégrer dans le futur SI. Cette dernière voie rejoint les propositions de (Mhiri, 2007) pour l'utilisation des ontologies comme moyen d'assistance à la conception des SI.

## 4. Enrichissement de l'ontologie de domaine.

L'enrichissement consiste à identifier de nouveaux éléments: concepts, termes et relations, puis de les placer dans une ontologie existante. L'enrichissement aussi bien que la construction manuelle d'une ontologie, s'avèrent être un travail fastidieux et coûteux (Ben Ghezaiel et al., 2010). Plusieurs travaux ont proposé des méthodes automatisées ou semi-automatisées d'enrichissement et de construction des ontologies. La plupart de ces méthodes s'appuient sur des sources externes à partir desquelles de nouvelles connaissances sémantiques sont détectées, évalués puis placées au sein de l'ontologie à enrichir. La méthode de mesure de la similarité sémantique (encadré 2), fait appel à un traitement d'enrichissement de l'ontologie de domaine, quand cette dernière ne contient pas de relation sémantique entre les concepts à aligner. Afin de mettre en œuvre ce traitement et de démontrer sa faisabilité, nous avons retenu, et ce à titre d'exemple, deux règles parmi les différentes règles relatives aux relations sémantiques :
- R1: Deux concepts sont similaires si leurs équivalents voisins sont similaires.

En effet, selon (Benslimane et al., 2008) deux concepts sont similaires si leurs sous-concepts "fils" sont similaires. Ceci a été confirmé dans (Fan et al., 2007).
- R2 : Deux concepts sont similaires si leurs sous-concepts "fils" sont similaires.

La règle R2 porte sur les concepts composites. Les concepts composites représentent les concepts père et les sous-concepts composants, liés par une relation sémantique de type part-of, sont les concepts fils.

Soient C1 et C2 les concepts à aligner et OCMi l'ontologie locale dont ils sont issus ; nous distinguons trois cas de figure :

**Cas n° 1** : C1 et C2 admettent une relation sémantique au sein de OCMi.
Cette relation est alors injectée dans l'ontologie de domaine OD.

**Cas n° 2** : C1 et C2 n'admettent pas de relation sémantique dans OCMi alors qu'il existe dans OCMi deux concepts C'1 et C'2 ainsi que deux relations sémantiques d'équivalence ; la première entre C1 et C'1 et la seconde entre C2 et C'2.

D'après R1 on peut déduire une nouvelle relation sémantique entre C1 et C2 qu'on injectera dans l'ontologie de domaine OD.

**Cas n° 3** : C1 et C2 sont des concepts composites qui n'admettent pas de relation sémantique dans OCMi, alors qu'il existe des relations sémantiques entre leurs sous concepts fils respectifs.

Soit {C11, C12, … C1n} l'ensemble des sous concepts fils de C1et {C21, C22,…. C2n} l'ensemble des sous concepts fils de C2, tel que C1i est C2i admettent une relation sémantique au sein de OCMi. D'après R2, on peut déduire une nouvelle relation sémantique entre C1 et C2 qu'on injectera dans l'ontologie de domaine OD.

### 5.   Prototype de validation

Afin d'évaluer et de valider la méthode, nous avons développé un prototype que nous avons baptisé IntegrateBusinessOnto (IBO). Ce prototype permettra de mettre en œuvre la méthode sur des exemples d'application.

Le prototype IBO peut être vu comme une extension du projet Growl (Krivov et al. 2007). Le choix de Growl peut être justifié par différentes raisons notamment :

- Growl permet la gestion des ontologies conformément à la nome W3C.
- Il est Open source.
- Il intègre l'API populaire Jena pour la gestion des ontologies.
- Il permet la visualisation des ontologies.
- Il est développé en JAVA, ce qui permet une interaction et intégration facile avec toute API développée en java pour la gestion des ontologies.

IBO supporte toutes les étapes du processus d'intégration. Il prend tout d'abord en entrée l'ontologie de domaine ainsi que les composants métier candidats à l'intégration. Il transforme ensuite Les CM fournis en ontologies décrites en OWL ; puis calcule les mesures de similarité et le cas échéant procède à l'enrichissement de l'ontologie de domaine. Enfin IBO délivre en sortie une description en OWL de l'ontologie résultat de l'intégration ainsi que sa représentation graphique.

Le prototype a été mis en œuvre pour l'intégration de composants métier relatifs au domaine médical.

## 6. Conclusion et perspectives.

Nous nous sommes intéressés dans ce travail à la résolution des conflits sémantiques de type nommage lors de la réutilisation des composants métiers conceptuels dans les phases d'analyse et de conception. Notre solution est basée sur une application des ontologies au domaine à la conception des SI par réutilisation des composants métier conceptuels. Elle consiste en un processus formé de trois étapes. La première et la dernière étape concernent la transformation des représentations conceptuelles des composants métier en représentations ontologiques et réciproquement. La deuxième étape, et qui constitue la partie fondamentale de notre travail, consiste en une méthode de calcul de la similarité sémantique avec enrichissement de l'ontologie de domaine. Nous avons également développé un prototype en vue de valider la solution. Les résultats obtenus par la mise en œuvre de notre prototype sur des exemples sont encourageants. Nous pensons poursuivre ce travail d'abord par une validation formelle de la solution, et ensuite par la recherche des possibilités de l'étendre pour résoudre les autres types de conflits sémantiques, notamment les conflits de mesure et de confusion.